\documentclass[usegraphicx,useAMS]{mn2e}

\usepackage{subfigure}
\usepackage{epsfig}
\usepackage{psfig}
\usepackage{amsfonts}
\usepackage{amsmath}
\usepackage{amssymb}
\usepackage{times}

\title[Candles with an intrinsic distribution]{Method of determining cosmological parameter ranges with samples of candles with an intrinsic distribution}
\author[Qin et al]{Y.-P. Qin$^{1.2}$\thanks{E-mail:
ypqin@ynao.ac.cn}, B.-B. Zhang$^{1,3}$,Y.-M. Dong$^{1,3}$, F.-W.
Zhang$^{2}$,\newauthor H.-Z. Li,$^{1,3}$, L.-W. Jia$^{1,3}$, L.-S.
Mao$^{1,3}$, R.-J. Lu$^{1,2,3}$,\newauthor T.-F. Yi $^{2}$, X.-H.
Cui$^{1,3}$, and Z.-B. Zhang
$^{1,3}$ \\
$^{1}$National Astronomical Observatories/Yunnan
Observatory, Chinese Academy of Sciences, P. O. Box 110, Kunming,\\
Yunnan, 650011, P. R. China\\
$^{2}$Physics Department, Guangxi University, Nanning,
Guangxi 530004, P. R. China\\
$^{3}$The Graduate School
of the Chinese Academy of Sciences}

\begin{document}

\date{Accepted year mon day . Received year mon day; in original form 2005,June,13th}

\pagerange{\pageref{firstpage}--\pageref{lastpage}} \pubyear{2002}

\maketitle

\label{firstpage}

\begin{abstract}
In this paper, the effect of the intrinsic distribution of
cosmological candles is investigated. We find that, in the case of
a narrow distribution, the deviation of the observed modulus of
sources from the expected central value could be estimated within
a ceratin range. We thus introduce a lower and upper limits of
$\chi ^{2}$, $\chi _{\min }^{2}$ and $ \chi _{\max }^{2}$, to
estimate cosmological parameters by applying the conventional
minimizing $\chi ^{2}$ method. We apply this method to a gamma-ray
burst (GRB) sample as well as to a combined sample including this
GRB sample and an SN Ia sample. Our analysis shows that: a) in the
case of assuming an intrinsic distribution of candles of the GRB
sample, the effect of the distribution is obvious and should not
be neglected; b) taking into account this effect would lead to a
poorer constraint of the cosmological parameter ranges. The
analysis suggests that in the attempt of constraining the
cosmological model with current GRB samples, the results tend to
be worse than what previously thought if the mentioned intrinsic
distribution does exist.
\end{abstract}


\section{Introduction}

One of the greatest achievements obtained in the past few years in
astrophysics is the determination of cosmological parameters with
type Ia supernovae (SN Ia), which suggests an accelerating
universe at large scales (Riess et al. 1998, Perlmutter et al.
1999, Tonry et al. 2003, Barris et al. 2004, Knop et al. 2003,
Riess et al. 2004). The cosmic acceleration was also confirmed,
independently of the SN Ia magnitude-redshift relation, by the
observations of the cosmic microwave background anisotropies
(WMAP: Bennett et al. 2003) and the large scale structure in the
distribution of galaxies (SDSS: Tegmark et al. 2004a, 2004b). It
is well known that all known types of matter with positive
pressure generate attractive forces and decelerate the expansion
of the universe. Given this, a dark energy component with negative
pressure was generally suggested to be the invisible fuel that
drives the current acceleration of the universe. There are a huge
number of candidates for the dark energy component in the
literature, such as a cosmological constant $\Lambda$ (Carroll et
al. 1992), an evolving scalar field (referred to by some as
quintessence: Ratra and Peebles 1988; Caldwell et al. 1998), the
phantom energy, in which the sum of the pressure and energy
density is negative (Caldwell 2002), the so-called ``X-matter"
(Turner and White 1997; Zhu 1998; Zhu, Fujimoto and Tatsumi 2001;
Zhu, Fujimoto and He 2004b), the Chaplygin gas (Kamenshchik et al.
2001; Bento et al. 2002; Zhu 2004), the Cardassion model (Freese
and Lewis 2002; Zhu and Fujimoto 2002, 2003, 2004; Zhu, Fujimoto
and He 2004a), and the brane world model (Randall and Sundrum
1999a, 1999b; Deffayet, Dvali and Gabadadze 2002).

Samples of SN Ia sources available in the early analysis contain only
sources with redshifts $z<1$. Although observations of the fluctuation in
the cosmic microwave background (CMB) can constrain the cosmological model
up to redshifts as high as $z\sim 1000$ (e.g., Spergel et al. 2003), a more
direct measurement of the universe with objects located at very large
distances is strongly desired. Fortunately, recent observations extended the
SN Ia sample to sources with redshifts as large as $z=1.7$. The previous
result was confirmed by these high redshift sources and the analysis
revealed that before its acceleration the universe underwent a period of
deceleration (Riess et al. 2004). The success of including high redshift SN
Ia sources inspires us to great efforts to search for cosmological rulers
with much higher redshifts. Based on the $E_{p}-E_{\gamma }$ relation found
recently in a class of gamma-ray bursts (GRBs) (Ghirlanda et al. 2004b), Dai
et al. (2004) assumed that the GRB sources obeying this relation can be used
to measure the universe. In their sample of 12 GRBs, two have redshifts $z>2$%
. Soon after their work, the same issue was investigated by many
authors (see Ghirlanda et al. 2004a; Friedman and Bloom 2005;
Firmani et al. 2005; Xu et al. 2005; Liang and Zhang 2005). It was
found that current GRB data which are lack of low redshift sources
could be used to marginalize some parameters in their reasonable
ranges (see Xu et al. 2005 for a detailed explanation), or they
could be employed to constrain the cosmological model with a new
Bayesian method (Firmani et al. 2005). Although the size of the
current GRB sample is small and low reshift sources are missed,
the idea that some high redshift extragalactic sources other than
SN Ia might be employed to determine the cosmological model is
quite interesting and promising.

It would be natural that, for a kind of source which could serve
as candles, one assumes a distribution of luminosity, which is
reasonable due to fluctuation. As discussed in Kim et al. (2004),
the uncertainty of a source must include both the systematic
uncertainty and the magnitude dispersion. We argue that, if there
exists a distribution of luminosity of the candles, the expected
luminosity itself (or the corresponding deduced luminosity
distance) could be different from source to source, which would be
due to an intrinsic property rather than to the measurement
uncertainty. This raises a topic of finding an appropriate method
to estimate cosmological parameter ranges with candles with a
certain distribution.

When employing candles such as SN Ia or GRBs to measure the
universe, the confidence level associated with the fit of the
theoretical curve to the luminosity distance data was described by
a statistic $\chi ^{2}$ which is defined under the assumption that
the measurement uncertainty is the only cause of the deviation of
the data to the curve. The best fit will be obtained when one
reaches the minimum value of $\chi ^{2}$. However, for candles
with a certain distribution, the deviation of the observed
luminosity from the expected curve must be caused by both the
measurement uncertainty and the distribution itself. When taking
into account the distribution of luminosity, the $\chi ^{2}$
statistic could not be defined if the distribution itself is
unknown. The minimizing $\chi ^{2}$ method will not be applicable
if the statistic itself cannot be defined.

In the following, we will study how to deal with this matter and
investigate what one can expect from the analysis. A corresponding
method will be proposed and will be illustrated with two samples.

\section{The method}

In this section, we propose a method to deal with candles with a
certain distribution when employing them to constrain the
cosmological model. As mentioned above, the statistic $\chi ^{2}$
could not be defined for candles with a certain distribution if
the distribution itself is unknown. Even if the distribution is
known, the statistic is still undefinable since there is no way to
know the real luminosity of each source. These difficulties lead
to two problems. One is that the well-known minimizing $\chi ^{2}$
method could not be applicable without a definition of the
statistic. The other is that the probability associated with the
statistic $ \chi ^{2}$, if we define it when taking into account
the deviation arising from the distribution, is not available
(since the real luminosity of each source is unknown).

It is known that the convolution of two Gaussian is still a
Gaussian with a width that is given by the quadratic sum of the
two widths of the original distributions. That is
$\sigma^2=\sigma_1^2+\sigma_2^2$, where $\sigma_1^2$ and
$\sigma_2^2$ are the variances of the two Gaussian functions
concerned and $\sigma^2$ is that of the resulted Gaussian.

Let us consider the deviation of an observed luminosity distance modulus, $%
\mu _{ob}$, of a source from the real value of the quantity, $\mu
_{th}$, which follows

\begin{eqnarray}
(\mu _{ob}\pm \sigma _{ob})-\mu _{th}(z;H_{0},\Omega _{m},\Omega
_{\Lambda }) \nonumber \\
=(\mu _{ob}\pm \sigma _{ob})-[\mu _{th,0}(z;H_{0},\Omega
_{m},\Omega _{\Lambda })+\Delta \mu _{th}],
\end{eqnarray}
where $\sigma _{ob}$ is the measurement uncertainty of $\mu
_{ob}$, $\mu _{th,0}$ is the central value of $\mu _{th}$, which
is the real value of the modulus expected in the case when there
is no distribution of the candles, and $\Delta \mu _{th}$
represents the deviation of $\mu _{th}$ from $\mu _{th,0}$.
Suppose that the distribution of candles is narrow enough so that
the absolute value of the deviation of $\mu _{th}$ from $\mu _{th,0}$, $%
|\Delta \mu _{th}|$, is small. According to the error transform
formula, the uncertainty of $\mu _{ob}$ relative to $\mu _{th,0}$
could be determined by
\begin{equation}
\sigma _{ob,0}=\sqrt{\sigma _{ob}^{2}+(\Delta \mu _{th})^{2}}.
\end{equation}%
Relative to the expected central moduli, the $\chi ^{2}$ statistic of\ a
sample of the candles could be determined by%
\begin{equation}
\chi ^{2}=\underset{i}{\sum }\frac{[\mu _{ob,i}-\mu
_{th,0,i}(z;H_{0},\Omega _{m},\Omega _{\Lambda })]^{2}}{\sigma
_{ob,i}^{2}+(\Delta \mu _{th,i})^{2}}.
\end{equation}%
[Note that, in the case of SN Ia, $\sigma _{ob,i}^{2}$ should be replaced by
$\sigma _{ob,i}^{2}+\sigma _{v}^{2}$, where $\sigma _{ob,i}$ is the
uncertainty in the individual distance moduli deduced from the empirical
relation between the light-curve shape and luminosity and $\sigma _{v}$ is
the uncertainty associated with the dispersion in supernovae redshift
(transformed to units of distance moduli) due to peculiar velocities (see
Riess et al. 2004)]

It seems that, with equation (3), one might be able to evaluate
the $\chi ^{2} $ statistic. But because $\Delta \mu _{th,i}$ is in
no way to be known, this is unfortunately not true. However, under
the condition that the distribution of candles is narrow, we can
estimate $\Delta \mu _{th,i}$ with the width of the distribution.
Let $\widetilde{\sigma }_{dis}$ be the width of the distribution
of $\mu _{th}/\mu _{th,0}$ (called the intrinsic distribution of
the relative luminosity distance moduli). (Note that $\mu
_{th}/\mu _{th,0}$ should of course become unity when there is no
deviation of $\mu _{th}$ from $\mu _{th,0}$). We assume $|\Delta
\mu _{th,i}|\simeq \widetilde{\sigma }_{dis}\mu _{th,0,i}$. Thus
the $\chi ^{2}$ statistic
could be estimated by%
\begin{equation}
\chi ^{2}\simeq \underset{i}{\sum }\frac{[\mu _{ob,i}-\mu
_{th,0,i}(z;H_{0},\Omega _{m},\Omega _{\Lambda })]^{2}}{\sigma _{ob,i}^{2}+%
\widetilde{\sigma }_{dis}^{2}\mu _{th,0,i}^{2}}.
\end{equation}

As long as $\widetilde{\sigma }_{dis}$ is provided, the $\chi
^{2}$ statistic is then available according to (4). For any kind
of candle, quantity $\widetilde{\sigma }_{dis}$ could be estimated
when the sample employed is large enough and the measurement
uncertainty $\sigma _{ob}$ is small enough and when the
cosmological model is fixed. Obviously, this could not be realized
at present since the cosmological model itself is currently a
target to be pursued and for interesting candles the measurement
uncertainty is always quite large. But this cannot prevent one to
estimate the limits of $\widetilde{\sigma }_{dis}$. As the
deviation of $\mu _{ob}$ from $\mu _{th,0}$ is caused by both the
distribution of $\mu _{th}$ and the measurement uncertainty of
$\mu _{ob}$ itself, $\widetilde{\sigma }_{dis}$
must be smaller than $\widetilde{\sigma }_{dis,\max }$, where $\widetilde{%
\sigma }_{dis,\max }$ is the width of the distribution of $\mu _{ob}/\mu
_{th,0}$, which is determined by $\widetilde{\sigma }_{dis,\max }=\sqrt{%
\underset{i}{\sum }(\mu _{ob,i}/\mu _{th,0,i}-1)^{2}/(N-1)}$, with $N$ being
the size of the sample. Let us over estimate the effect of the measurement
uncertainty in the opposite way. Within the range of $[\mu _{ob,i}-\sigma
_{ob,i},\mu _{ob,i}+\sigma _{ob,i}]$ we take the value that is the closest
one to $\mu _{th,0,i}$ as $\mu _{ob,i}^{\ast }$. Obviously, the distribution
of $\mu _{ob}^{\ast }/\mu _{th,0}$ would be narrower than the distribution
of $\mu _{th}/\mu _{th,0}$ since the deviation caused by the measurement
uncertainty is over subtracted. We take the width of the distribution of $%
\mu _{ob}^{\ast }/\mu _{th,0}$ as $\widetilde{\sigma }_{dis,\min }$, which
is calculated with $\widetilde{\sigma }_{dis,\min }=\sqrt{\underset{i}{\sum }%
(\mu _{ob,i}^{\ast }/\mu _{th,0,i}-1)^{2}/(N-1)}$. Clearly, $\widetilde{%
\sigma }_{dis}$ must be larger than $\widetilde{\sigma }_{dis,\min }$. With
these two quantities we have%
\begin{equation}
\chi _{\min }^{2}\simeq \underset{i}{\sum }\frac{[\mu _{ob,i}-\mu
_{th,0,i}(z;H_{0},\Omega _{m},\Omega _{\Lambda })]^{2}}{\sigma _{ob,i}^{2}+%
\widetilde{\sigma }_{dis,\max }^{2}\mu _{th,0,i}^{2}}
\end{equation}%
and%
\begin{equation}
\chi _{\max }^{2}\simeq \underset{i}{\sum }\frac{[\mu _{ob,i}-\mu
_{th,0,i}(z;H_{0},\Omega _{m},\Omega _{\Lambda })]^{2}}{\sigma _{ob,i}^{2}+%
\widetilde{\sigma }_{dis,\min }^{2}\mu _{th,0,i}^{2}}.
\end{equation}%
Since $\widetilde{\sigma }_{dis,\min }<\widetilde{\sigma }_{dis}<\widetilde{%
\sigma }_{dis,\max }$, one gets $\chi _{\min }^{2}<\chi ^{2}<\chi
_{\max }^{2}$. With equations (5) and (6), one can calculate the
corresponding probability associated with the $\chi ^{2}$
statistic and confine the conventional confidence contour. In this
way, cosmological parameters would be constrained. With this
estimating method, the first problem is largely eased and the
second is solved.

\section{Application}

Let us consider a GRB sample. The sample was presented and studied
in Xu et al. (2005) and Xu (2005) (the XDL GRB sample) which
contains 17 GRBs. As suggested in Ghirlanda et al. (2004a), the
scatter of the data points of their GRB sample around the
correlation of $E_{p}-E_{\gamma }$ found recently (Ghirlanda et
al. 2004b) is of a very small order.

To check if the data of the XDL GRB sample are consistent with no
scatter beyond the measurement errors in terms of statistics, the
simplest method is to calculate the mean of the deviation of the
deduced luminosity distance moduli from the expected one of the
sample and then compare it with the average of the measurement
error. The mean of the deviation is defined as
$\sigma_{dev}=\sqrt{%
\underset{i}{\sum }((\mu _{ob,i}-\mu _{ex,i})/\mu
_{ex,i})^{2}/(N-1)}$, where $\mu _{ex}$ is the expected value of
$\mu$, while the average of the measurement error is calculated
with
$\sigma_{err}=\sqrt{%
\underset{i}{\sum }(\sigma_{ob,i}/\mu _{ex,i})^{2}/(N-1)}$. (Note
that, as redshifts of these sources are not the same, we consider
the relative values.) We get the following from the XDL sample:
$\sigma_{dev}=0.0122$ and $\sigma_{err}=0.0116$, where we adopt
$(\Omega _{m},\Omega _{\Lambda },h)=(0.29,0.71,0.65)$. It shows
that the deviation is slightly larger than the measurement error.
(Ignoring the slight difference between the two quantities, the
result confirms what suggested in Ghirlanda et al. 2004a, 2004b.)
Taking $\mu _{th,0}$ as $\mu _{ex}$ adopted here, one finds that
$\sigma_{dev}$ is identical with $\widetilde{\sigma }_{dis,\max }$
defined in last section. Thus, for the XDL sample,
$\widetilde{\sigma }_{dis}<0.0122$, suggesting that the
distribution, if exists, would be quite narrow. Another approach
involves a simulation analysis. We assume that there is no
intrinsic distribution of the deduced luminosity distance moduli,
and thus the deviation observed is due to the measurement
uncertainty. Obviously, under this assumption the distribution of
$\mu _{ob}/\mu _{ex}$ should peak at unity. According to the null
hypothesis, the observed value of $\mu _{ex}$ for each source is
obtained by chance from a parent population of $\mu _{ob}^{\prime
}$ whose distribution obeys a Gaussian with the measurement
uncertainty served as the width of the Gaussian. For each source
one can create a $\mu _{ob}^{\prime }$ via simulation as long as
the expected value $\mu _{ex}$ and the measurement uncertainty are
known. In this way, from the 17 $\mu _{ex}$ and the corresponding
measurement uncertainties, one can create a set of 17 $\mu
_{ob}^{\prime }$ data by a Monte-Carlo simulation and then obtain
a set of 17 $\mu _{ob}^{\prime }/\mu _{ex}$ data. We perform 100
times of simulation and get 100 sets of 17 $\mu _{ob}^{\prime
}/\mu _{ex}$ data. Combining these 100 sets we get a large sample
with its size being 1700. The deviation of the relative simulated
luminosity distance moduli from the expected one (the unity) is
defined as
$\sigma_{dev}^{\prime }=\sqrt{%
\underset{i}{\sum }(\mu _{ob,i}^{\prime }/\mu
_{ex,i}-1)^{2}/(N-1)}$. Note that $\sigma_{dev}^{\prime }$ could
be written as
$\sigma_{dev}^{\prime }=\sqrt{%
\underset{i}{\sum }((\mu _{ob,i}^{\prime }-\mu _{ex,i})/\mu
_{ex,i})^{2}/(N-1)}$, which could thus be directly compared with
$\sigma_{dev}$, the deviation of the observed data defined above.
From the XDL sample we get $\sigma_{dev}^{\prime }=0.0113$, which
suggests that the deviation associated with observation, denoted
by $\sigma_{dev}$, is also slightly larger than that expected from
the measurement uncertainties. Two methods come to almost the same
result, suggesting that there might be an intrinsic distribution
of the relative luminosity distance moduli of the XDL sample,
although it would be quite narrow (as the difference between
$\sigma_{dev}^{\prime }$ and $\sigma_{dev}$ and that between
$\sigma_{err}$ and $\sigma_{dev}$ are small).

To illustrate how to apply the method proposed above to deal with
data with intrinsic distributions, we assume in the following that
there is a distribution of the true value of the deduced relative
luminosity distance moduli for the XDL sample, although the
distribution, if it exists, might be very narrow (see what
suggested above). For the sake of comparison, we perform the fit
with three $\chi ^{2}$ statistics. One is the conventional $\chi
^{2}$ which could be determined by (3) when taking $\Delta \mu
_{th,i}=0$. The other two are $\chi _{\min }^{2}$ and $\chi _{\max
}^{2}$ which are determined by equations (5) and (6) respectively.
Each $\chi ^{2}$ statistic is calculated with the XDL GRB sample
in many tries. In each try, we adopt a set of parameters and based
on these parameters we deduce both the observed and theoretical
luminosity distance moduli. With these moduli and the measurement
uncertainties, we are able to evaluate $\widetilde{\sigma
}_{dis,\min }$ and $\widetilde{\sigma }_{dis,\max }$ (see what
proposed in last section), and then the corresponding $\chi ^{2}$
statistic would be well determined ($H_{0} = 65 kms^{-1}Mpc^{-1}$
is adopted throughout this paper). For each $\chi ^{2}$, the best
fit will be obtained when the smallest value is reached.

Displayed in Fig. 1 are the Hubble diagram and the confident
contour plot of the XDL GRB sample. As concluded previously by
other authors (see Ghirlanda et al. 2004a; Friedman and Bloom
2005; Xu et al. 2005), currently, employing GRB samples alone
cannot tightly constrain the cosmological model. Fig. 1 shows
that, the parameter ranges are indeed poorly constrained even
there is no intrinsic distribution of the relative luminosity
distance moduli (see solid lines in Fig. 1b). Taking into account
an intrinsic distribution of the moduli leads to much poorer
results. This indicates that if there indeed exists an intrinsic
distribution of the moduli, the effect arising from the
distribution should not be ignored.

Shown in Table 1 are the best fit cosmological parameters for the
three kinds of universe, obtained by applying the minimizing $\chi
^{2}$ method to the three $\chi ^{2}$ statistics, where the
$1\sigma$ errors are estimated from the corresponding $1\sigma$
contours in Fig. 1b. As shown in Fig. 1b, the $1\sigma$ contours
are not closed within the ranges of the plot. This leads a poor
constraint to the limits of the parameters. Some limits are
therefore not able to be determined, which are denoted by ``?'' in
Table 1.

\begin{figure}
  \centering
  \subfigure[]{
    \label{fig:subfig:a} 
    \includegraphics[width=3.0in]{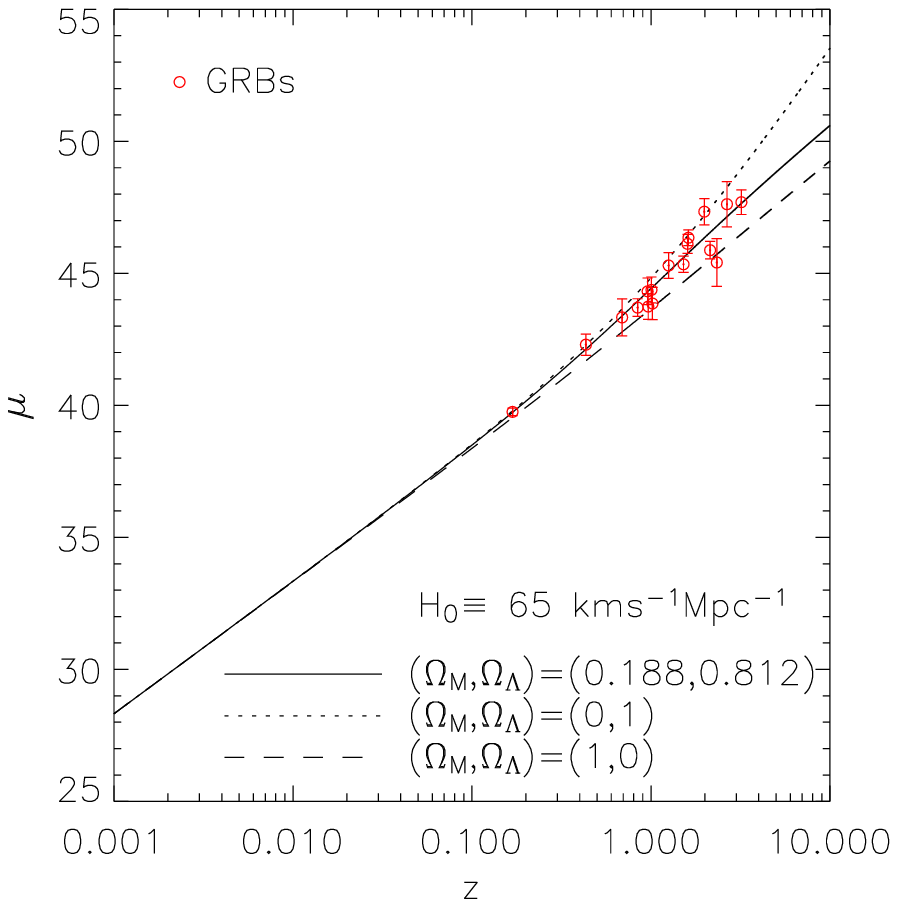}}
  \subfigure[]{
    \label{fig:subfig:b} 
    \includegraphics[width=3.0in]{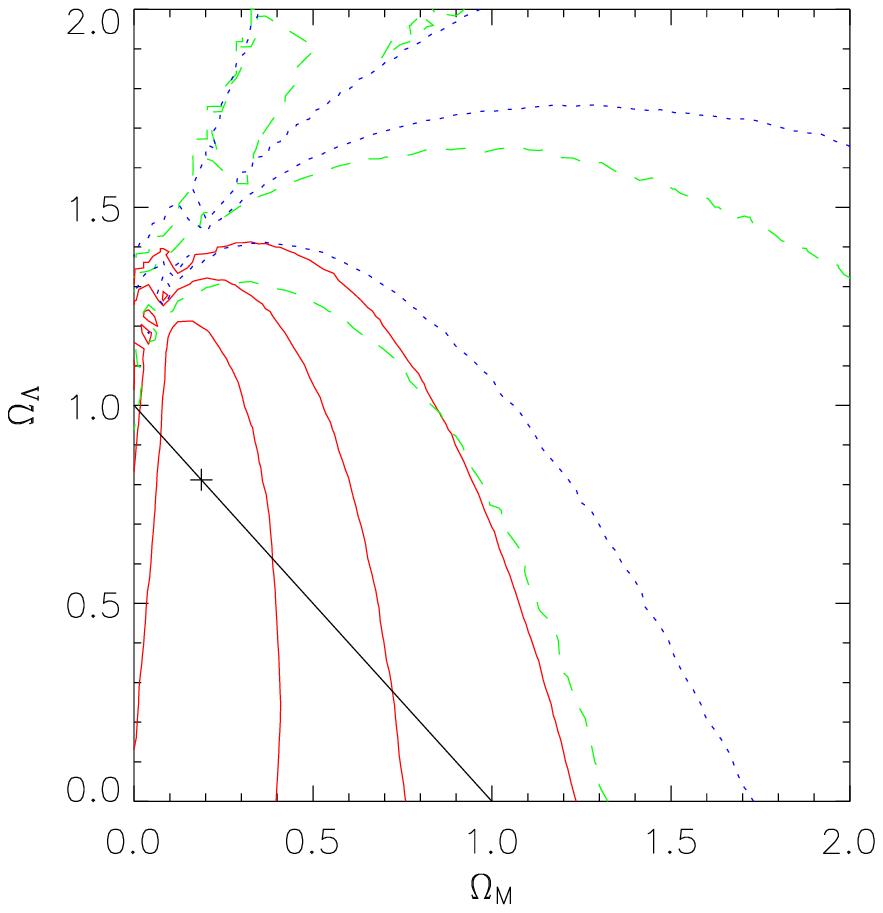}}
   \caption{(a) Hubble diagram of the XDL GRB sample, where the two edge curves (the dotted
and dashed lines) of the flat universe are also represented. (b)
Confident contour plot for the XDL GRB sample, where three dashed
lines from the innermost curves to the outmost one represent the
1, 2, and 3 $\sigma $\ levels of confidence calculated with the
statistic $ \chi _{\min }^{2}$\, respectively, while the three
dotted lines represent those associated with the statistic $\chi
_{\max }^{2} $\, respectively. For the sake of comparison, the
confidence levels calculated without considering the distribution of $%
\widetilde{\mu }_{ob}/\mu _{th}$\ are also plotted (the solid
lines). The straight line denotes the flat universe and the plus
represents the best fit parameters of the flat universe obtained
by the conventional minimizing $\chi ^{2}$ method.}
\end{figure}

\begin{table*}
 \centering
\begin{minipage}{150mm}
  \caption{Best-fit parameters obtained by the least square method}
\begin{tiny}
\begin{tabular}{lllll}
\hline
\hline
Sample&Universe & $(\Omega_{M},\Omega_{\Lambda},\chi_{0,\nu}^2)$ \footnote{$\chi_{0,\nu}^2$ is the reduced $\chi^2$ calculated with equation (4) when assigning $\widetilde{\sigma }_{dis}=0$.}& $(\Omega_{M},\Omega_{\Lambda},\chi_{max,\nu}^2)$ &$(\Omega_{M},\Omega_{\Lambda},\chi_{min,\nu}^2)$ \\
\hline
SN+GRB & flat & $(0.283^{+0.0314}_{-0.0288},0.717, 197.9)$ &$(0.288^{+0.0134}_{-0.0201},0.712, 193.1)$ & $(0.288^{+0.0138}_{-0.000375},0.712, 185.4)$ \\ 
\hline
SN+GRB & open & $(0.368^{+0.127}_{-0.114},0.857^{+0.371}_{-0.170}, 197.0)$ & $(0.281^{+0.0201}_{-0.0201},0.717^{+0.0198}_{-0.0353}, 193.2)$ & $(0.281^{+0.00669}_{-0.0134},0.717^{+0.0100}_{-0.0201}, 185.4)$ \\ 
\hline
SN+GRB & closed & $(0.281^{+0.0334}_{-0.0469},0.717^{+0.0296}_{-0.0.0804}, 198.0)$ & $(0.428^{+0.147}_{-0.161},0.942^{+0.226}_{-0.246}, 191.3)$ & $(0.441^{+0.147}_{-0.0201},0.967^{+0.226}_{-0.0351}, 183.1)$ \\ 
\hline
GRB & flat & $(0.188^{+0.200}_{-0.114},0.812, 19.69)$  &  $(0.188^{+1.176}_{-?},0.812, 15.14)$ & $(0.188^{+1.539}_{-?},0.812, 7.43)$ \\ 
\hline
GRB & open & $(0.187^{+0.221}_{-?},0.682^{+0.221}_{-?}, 19.67)$  & $(0.187^{+?}_{-?},0.256^{+?}_{-?}, 14.96)$  & $(0.154^{+?}_{-?},0.391^{+?}_{-?}, 7.40)$ \\ 
\hline
GRB & closed &$(0.187^{+0.201}_{-0.114},0.817^{+0.401}_{-0.206}, 19.67)$ & $(0.187^{+1.177}_{-?},0.817^{+0.551}_{-?}, 15.14)$  &  $(0.187^{+1.54}_{-?},0.817^{+0.59}_{-?}, 7.44)$ \\
\hline
\hline
\end{tabular}
\end{tiny}
\end{minipage}
\end{table*}

The fact that the parameter ranges are poorly constrained (even
when the intrinsic distribution of the relative luminosity
distance moduli is ignored) might probably be due to the lack of
low redhsift sources, as it is already known that low redhsift
sources are important when employing a GRB sample to constrain the
cosmological parameters (see Firmani et al. 2005). We thus follow
what were done previously (see Ghirlanda et al. 2004a) to combine
an SN Ia sample and the XDL sample to constrain the cosmological
model. The SN Ia sample employed is that presented in Riess et al.
(2004) (the so-called gold set of SN Ia) which contains 157
sources (where, many low redshifts sources are included). In the
same way and for the same reason we apply the minimizing $\chi
^{2}$ method to the three $\chi ^{2}$ statistics to find the best
fit cosmological parameters. Note that, unlike what is shown in
the case of the GRB sample, the deduced luminosity distance moduli
of the SN\ Ia sources do not depend on the adopted cosmological
parameters.

It is known that, in estimating the deduced luminosity distance
moduli of the SN\ Ia sources, deviations caused by different
magnitudes of the peak luminosity of the sources have been
checked. Indeed, we find that the distribution of the relative
luminosity distance moduli of the SN Ia sample is very narrow (the
figure is omitted). This suggests that, if it still exists
(possibly caused by the small deviation from the adopted empirical
relation between the light-curve shape and luminosity), the
intrinsic distribution must be extremely narrow. Thus we ignore
the intrinsic distribution of the relative luminosity distance
moduli of the SN Ia sample, and consider only $\widetilde{\sigma
}_{dis,\min }$ and $\widetilde{\sigma }_{dis,\max }$ for the GRB
sample when we calculate the corresponding $\chi _{\min }^{2}$ and
$\chi _{\max }^{2}$ for the combined sample (including the XDL GRB
sample and the gold SN Ia sample).

Shown in Fig. 2 are the Hubble diagram and the confident contour
plot of the combined sample. One finds that, including the SN Ia
sample significantly improves the constraint of the ranges of
cosmological parameters. Once more, the result shows that taking
into account the intrinsic distribution of the relative luminosity
distance moduli leads to a poorer constraint. The effect is still
obvious (although it is less obvious than that adopting the GRB
sample alone) and therefore should not be neglected.
\begin{figure}
  \centering
  \subfigure[]{
    \label{fig:subfig:a} 
    \includegraphics[width=3.0in]{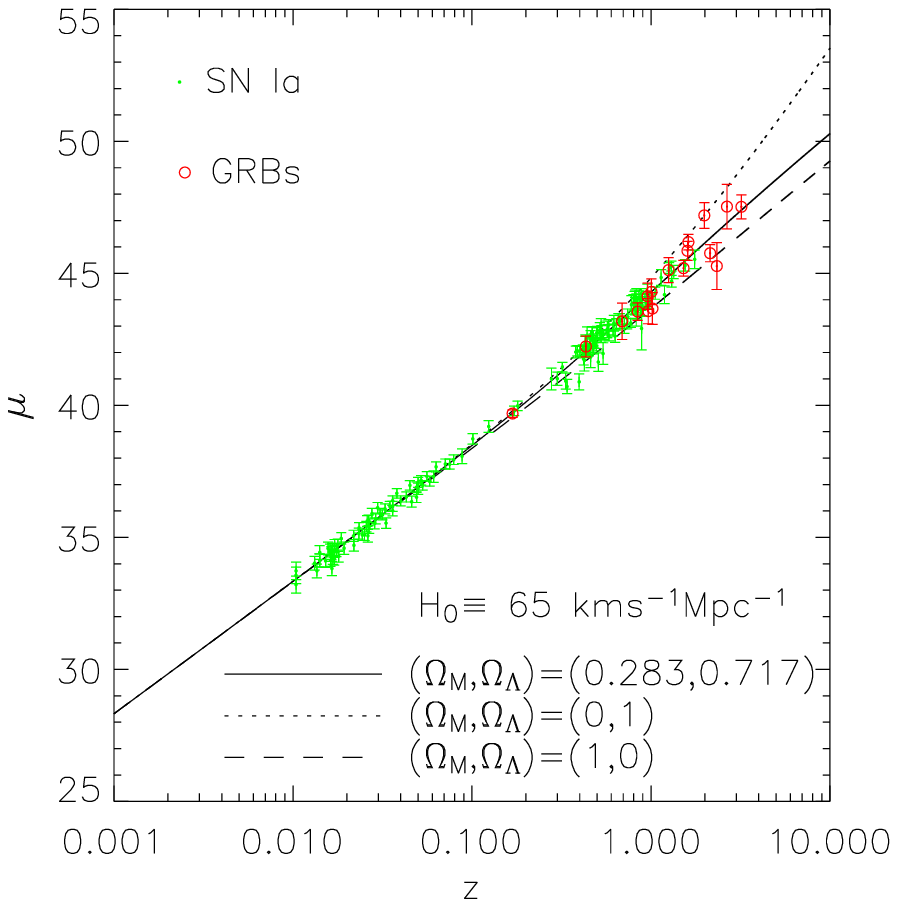}}
  \subfigure[]{
    \label{fig:subfig:b} 
    \includegraphics[width=3.0in]{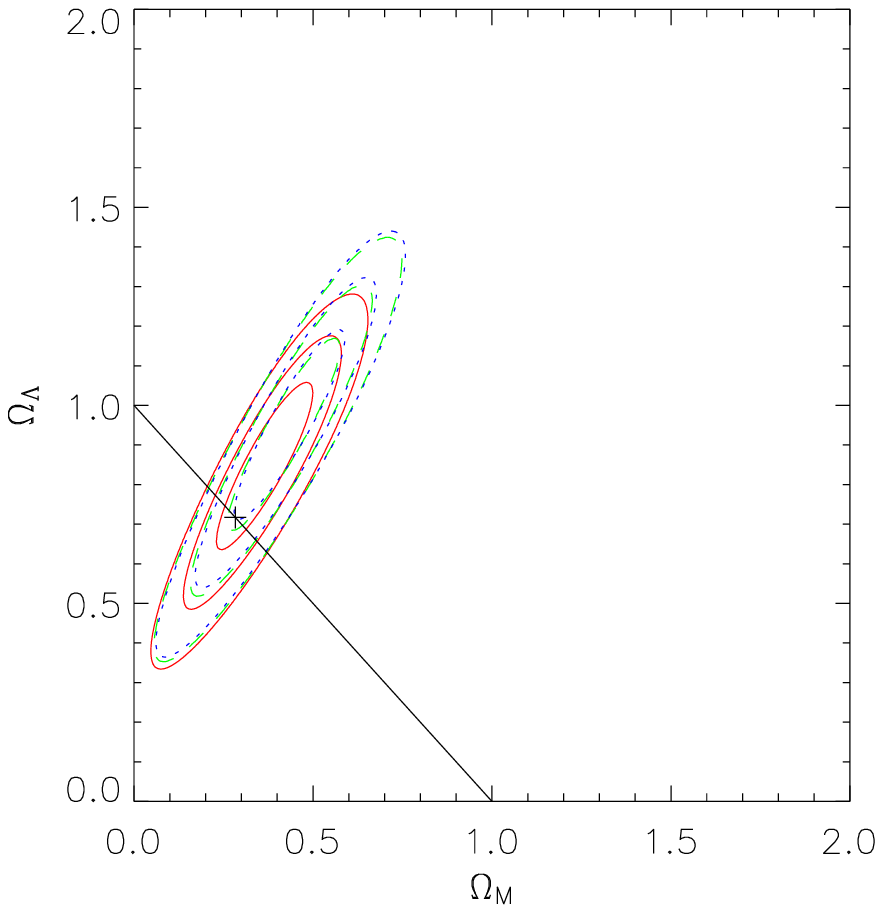}}
   \caption{(a) Hubble diagram of the combined sample, where the empty circle
   represents the XDL GRB sample and the filled circle stands for the gold SN Ia sample.
   Symbols of lines are the same as those denoted in panel (a) of Fig. 1. (b) Confident contour plot for the combined
sample, where the symbols are the same as those denoted in panel
(b) of Fig. 1.}
\end{figure}

The resulting best fit cosmological parameters as well as their
$1\sigma$ errors are listed in Table 1 as well.

\section{Discussion and conclusions}

The effect of the intrinsic distribution of cosmological candles
is investigated in this paper. Due to fluctuation, it is natural
that a property (say, the luminosity) of sources served as a
cosmological candle might form a distribution and scatter around a
central value. If the distribution does exist, the statistic $\chi
^{2}$ cannot be defined since the distribution itself is unclear
and the real value of the property for each source is unknown.
However, when the distribution is narrow, the deviation of the
observed modulus of each source from the central value could be
estimated within a ceratin range. We accordingly define a lower
and upper limits of $\chi ^{2}$, $\chi _{\min }^{2}$ and $ \chi
_{\max }^{2}$, to estimate cosmological parameters via the
conventional minimizing $\chi ^{2}$ method. The confidence
contours of these two $\chi ^{2}$ statistics can then be plotted
in the conventional way, and with these curves the ranges of the
parameters could be determined as long as a confidence level is
assigned.

With this method, a sample bearing a relatively small width of the
intrinsic distribution of the deduced relative luminosity distance
moduli would be applicable to constraining the cosmological
parameters. To illustrate this method we employ a GRB sample alone
and later combine this GRB sample with the gold SN Ia sample,
assuming that this GRB sample (the XDL sample) has an intrinsic
distribution of the deduced relative luminosity distance moduli
while the SN Ia sample has not. The analysis suggests that: a) the
effect of the intrinsic distribution of the relative luminosity
distance moduli is obvious and therefore should not be neglected
if the distribution itself does exist; b) taking into account this
effect would lead to a poorer constraint of the ranges of
cosmological parameters. This indicates that in the attempt of
constraining the cosmological model with GRB samples, the results
tend to be worse than what previously thought if the mentioned
intrinsic distribution exists, although the distribution is very
narrow.

As revealed recently by Wang et al. (2005), there is a clear
evidence for a tight linear correlation between peak luminosities
of SN Ia and their $B-V$ colors at $\sim $ 12 days after the $B$
maximum. They found that this empirical correlation allows one to
reduce scatters in estimating their peak
luminosities from $\sim $ 0.5 mag to the levels of 0.18 and 0.12 mag in the $%
V$ and $I$ bands, respectively. We wonder if taking into account
this effect can reduce the measurement uncertainty of the
luminosity distance of the SN Ia sources. If so, the ranges of the
cosmological parameters might be better constrained (when compared
with Fig. 2) (this will be investigated later).

As encountered in other cases, our method suffers from possible evolution of
candles. Quite recently, Firmani et al. (2004) found evidence supporting an
evolving luminosity function of long GRBs, where the luminosity scales as $%
(1+z)^{1.0\pm0.2}$. It is unclear if the corrected gamma-ray
energy, from which the luminosity distance moduli of the adopted
GRB sample are deduced, evolves with redshif. If so, the question
if the GRB sample can still be used to constrain the cosmological
model should be answered. This deserves a detailed investigation.
(It could be done only when the size of the sample is large
enough).

\section*{Acknowledgments}

We thank Profs. K. S. Cheng, C. Firmani, Z. G. Dai, Y.-Q. Lou, Y.-F. Huang,
and Z.-H. Zhu for their helpful suggestions and comments. This work was
supported by the Special Funds for Major State Basic Research Projects
(\textquotedblleft 973\textquotedblright ) and National Natural Science
Foundation of China (No. 10273019).

\bsp

\label{lastpage}

\end{document}